\documentstyle[aps,pre,epsf,twocolumn]{revtex}
\begin{document}
\draft
\parskip 1mm

\title  {Critical behaviour of annihilating random walk of two species
with exclusion in one dimension}

\author{G\'eza \'Odor}
\address{Research Institute for Technical Physics and Materials Science, \\
H-1525 Budapest, P.O.Box 49, Hungary}
      
\author{N\'ora Menyh\'{a}rd}
\address{Research Institute for Solid State Physics and Optics, \\
H-1525 Budapest,P.O.Box 49, Hungary}

\maketitle

\begin{abstract}
The $A+A\to 0$, $B+B\to 0 $ process with exclusion between the different
kinds is investigated here numerically. Before treating this model
explicitly, we study 
the generalized Domany-Kinzel cellular automaton model of Hinrichsen
on the line of the parameter space where only
compact clusters can grow. The simplest version is treated with
two absorbing phases in addition to the active one. The two kinds of
kinks which arise in this case do not react, leading to kinetics
differing from standard annihilating random walk of two species.
Time dependent simulations are presented here to illustrate the
differences caused by exclusion in the scaling properties
of usually discussed characteristic quantities. The dependence on the
density and composition of the initial state is most apparent.
Making use of the parallelism between this process
and directed percolation limited by a reflecting parabolic
surface we argue that the two kinds of kinks
exert marginal perturbation on each other leading to deviations
from standard annihilating random walk behavior.
\end{abstract}

\section{Introduction}

Non-universal dynamical behaviour seems to be a controversial
issue  in non-equilibrium models. An outstanding
example is the debated behaviour of systems exhibiting infinitely many
absorbing states \cite{PCP,GEP,Munoz,Mendes-Marcques}.
There is no analytic treatment up to now; argumentation
of various authors, in most of the cases, is based on 
simulation results. Despite intensive study,
the critical behaviour of such systems is poorly
understood, non-universality remains an unresolved problem
and even scaling behaviour is questioned.
Roughly speaking, in these coupled processes the 'primary' particles
follow a branching and annihilating random walk while the other
species just  provide a slowly changing environment that effects the
branching rates of the primaries. The spreading exponents of the
primaries depend on the initial conditions of the environment.

A possible way which might lead to a deeper understanding of the
mechanism behind non-universal spreading
could be the study of simpler coupled systems. Perhaps
the simplest case is the coupled annihilating random walk
 of two species ( $A+A\to\emptyset$, $B+B\to\emptyset$).
Naively one would expect that this could be described by the 
exactly solved field theory of the $A+A\to\emptyset$ 
process \cite{Lee} (ARW).
In one dimension, however, the situation is more subtle than in
higher  dimensions. Particles of different type can block the motion 
of each other. The difference between one and two dimensions has 
been found to give rise to different phase
diagrams in the case of the general epidemic model \cite{GEP}.
The question now arises how relevant the exclusion perturbation caused
by this blocking mechanism is to a fixed point of the kind
determined in \cite{Lee} .

An other motivation of this study originates from the investigations of
Hinrichsen \cite{GDK}, who found, by simulations, a strange scaling behaviour
in some special case of his model \cite{Hpriv},for which, however,
an explanation is still lacking.
In  section \ref{def} Hinrichsen's model will be introduced.
It is easy to see, that the kinks in this model at the symmetry 
point corresponding to the compact directed percolation point of the 
Domany-Kinzel automaton, exhibit the process described above.
In sections \ref{decaysim} and \ref{seedsim}
we present our high precision time dependent simulation results from
random and seed initial conditions. In section \ref{parabsim} 
these results are compared with those obtained by rigid (i.e. parabolic)
boundaries. 
We further investigate this analogy on the mean-field level in section
\ref{parabMF}, while  section \ref{ARW2e-sect} is devoted to results in the
explicit two-species Annihilating Random Walk model with exclusion (ARW2e).
We summarise our numerical results in section \ref{summary} and give an
outlook toward $N$-species generalisation in \ref{Gener}. A qualitative
description of the behaviour of Hinrichsen's model outside the
symmetry point on the line of compactness is presented
 in section \ref{pneq} and finally in section \ref{DISCUS} we summarise and
 discuss our results.

\section{The Generalised Domany-Kinzel SCA} \label{def}

The Domany-Kinzel (DK) stochastic cellular automaton (SCA)
\cite{DoKi} is one of the simplest models
which show
a non-equilibrium phase transition into an absorbing state.
This one dimensional SCA is defined on a ring with two states
'1' and '0'  with the following rule of update:
\begin{verbatim}
t:        0   0   0   1   1   0   1   1
t+1:        0       p       p       q

\end{verbatim}
where at $t+1$ the probability of '1'-s is shown .

In the plane of the parameters $(p,q)$, the phase diagram of the DK SCA
is as follows. A line of critical points separates the active phase
( with a finite concentration of '1'-s) from the absorbing (vacuum) 
phase (with zero steady state density of '1'-s). This continuous 
transition belongs to the universality class of directed percolation 
(DP) \cite{DP}.
The endpoint of this line ($q=1$, $p=1/2$) describes a transition,
however, outside the DP class; it corresponds to compact directed
percolation. Here the model exhibits Ising symmetry and can be solved 
exactly \cite{DoKi}.
\begin{center}
\begin{figure}[h]
\epsfxsize=80mm
\epsffile{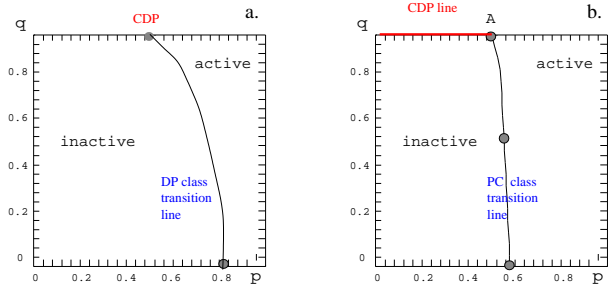}
\vspace{4mm}
\caption{a).The phase diagram of the DK SCA and b). the corresponding
phase diagram in case of the simplest version of the GDK cellular automaton
model by Hinrichsen}
\label{DKSCA}
\end{figure}
\end{center}

In 1997 Hinrichsen \cite{GDK} introduced a generalized version of the
DK model  including more than one symmetric inactive states
($I1$, $I2$, ...) and one active state ($A$). The motivation for this study
was to look for  a possible change in the universality class of the line
separating the
active and passive steady states. This generalized DK model
(GDK in the following),
in its simplest form with two absorbing states $I1$ and $I2$
has been defined by the rules given below:
\begin{center}
\begin{tabular}{||c||c|c|c||}
\hline
$s_{1},s_{2}$ & $P(A  \,|\,s_{1},s_{2})$ &
$P(I_1\,|\,s_{1},s_{2})$ & $P(I_2\,|\,s_{1},s_{2})$ \\ \hline\hline
$AA$     & $q$ & $(1-q)/2$ & $(1-q)/2$ \\ \hline
$AI_1$   & $p$ & $1-p$   & $0  $   \\ \hline
$AI_2$   & $p$ & $0$     & $1-p$   \\ \hline
$I_1A$   & $p$ & $1-p$   & $0$     \\ \hline
$I_2A$   & $p$ & $0$     & $1-p$   \\ \hline
$I_1I_1$ & $0$ & $1$     & $0$     \\ \hline
$I_1I_2$ & $1$ & $0$     & $0$     \\ \hline
$I_2I_1$ & $1$ & $0$     & $0$     \\ \hline
$I_2I_2$ & $0$ & $0$     & $1$     \\ \hline
\end{tabular}
\end{center}
The geometry of updating is the same as in the case of the DK SCA.
It has been shown by simulation \cite {GDK} that the phase diagram 
which emerges is similar to that in the DK SCA : an active phase is 
separated from an inactive one by a line of continuous phase 
transitions.
The inactive phase, however is symmetrically degenerated ($I1$ or $I2$)
and the phase transition line now belongs to the parity conserving
(PC) universality class. This class has been studied by many authors
as the first exception from the robust DP class
\cite{gra84,gra89,Taka,jen94,men94,MeOd95,MeOd96,Card97,Bassler}.

\begin{figure}[h]
\epsfxsize=70mm
\centerline{\epsffile{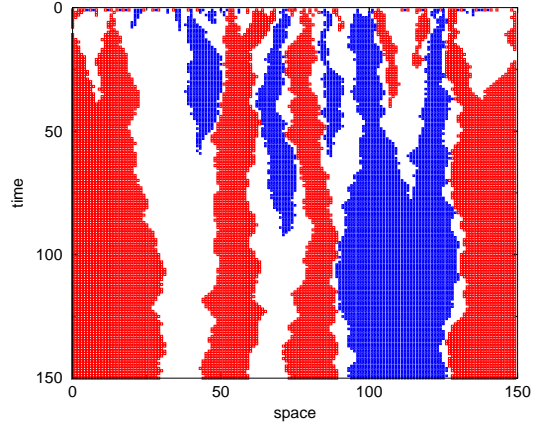}}
\caption{Evolution from a random initial state
in the generalised Domany-Kinzel SCA (GDK) on the line of compactness,
at $q=1$, $p=0.5$. Red: $I1$, blue: $I2$, white: $A$
\label{rajz_45}
}
\end{figure}

The phase diagram exhibited on Fig1.b) shows that the line of
PC transitions ends at $q=1$, $p=1/2$, a point which
corresponds the  Ising symmetry point of the DK automaton.
The primary aim of the present work is to investigate the scaling properties
of GDK at this point, which will be called CDP2 transition point.
A typical  time evolution of the
GDK model at this special point when starting from a random initial
arrangement of $I1$-s, $I2$-s and $A$-s is shown on   Fig. \ref{rajz_45}.
Here active islands can be spatially extended, thus three kinds of
compact clusters can grow. Nevertheless only the $I1$ and $I2$ phases
are $Z_2$-symmetric while the active phase plays a special role.
(The situation is different from a 3-states Potts model with Glauber
kinetics ).

It is well known that the CDP process in 1d is equivalent to an
annihilating random walk process of kinks \cite {DoKi}
separating compact domains of $0$-s and $1$-s.
In the  model investigated here two types of kinks can be defined,
 namely kink $K1$ between domains  $A-I1$ (and $I1-A$) and kink $K2$
between neighbouring $A-I2$-s (and $I2-A$-s). The rules of the
model inhibit occurrence of kinks between domains of absorbing
phases, i.e. between $I1$-$I1$ and $I2$-$I2$.

Kinks $K1$ and $K2$ perform annihilating random walks:  
$K1 + K1\to\emptyset$, $K2 + K2\to\emptyset$, while the processes 
$K1 + K2\to\emptyset$, $K2 + K1\to\emptyset$ are, however, forbidden. 
In other words, upon meeting, a  $K1$ and a $K2$  ``block''
each other (do not annihilate and do not exchange sites)\cite{Hpriv}.
To our knowledge such kind of kinetics has not been studied before.
Motivated by this fact we have decided to explore the critical 
behaviour of the above described system, on the line of compactness 
($q=1$), by computer simulation.
In this study special attention will be paid to  the
$p=1/2$ symmetry point CDP2.

\section{Simulations from random initial state} \label{decaysim}

\subsection{Kink decay simulations}

We have performed time dependent simulations starting from states
with uniformly distributed species $A$, $I1$, and $I2$, 
with respective densities:
$\rho_0(A)$,
$\rho_0(I1)$
and $\rho_0(I2)$. At the CDP2 point  unusual scaling behaviour of
the density of kinks has been  observed previously \cite{Hpriv}: a
deviation from the ordinary annihilation-diffusion process with
kink-density decay
$\rho(t)\sim\frac{1}{\sqrt{t}}$. Instead,
$\rho(t)\sim t^{-\alpha}$ with
$\alpha\approx 0.55$ has resulted from the first simulations.

\begin{figure}[h]
\epsfxsize=70mm
\centerline{\epsffile{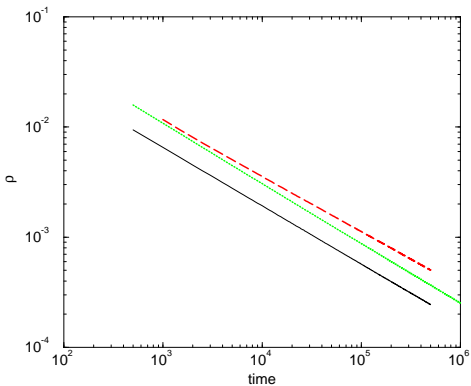}}
\caption{Total kink number as the function of time started from 
symmetrical homogeneous random initial states: 
$\rho_0(I1)=\rho_0(I2)=0.3$ (solid line) and 
$\rho_0(I1)=\rho_0(I2)=0.1$ (dotted line).
The dashed line corresponds to the single species annihilating 
random walk ($\rho_0(I1)=0$, $\rho_0(I2)=1/2$) 
exhibiting $\rho(t) \propto t^{-0.5}$.
\label{decay}
}
\end{figure}

\begin{figure}[h]
\epsfxsize=70mm
\centerline{\epsffile{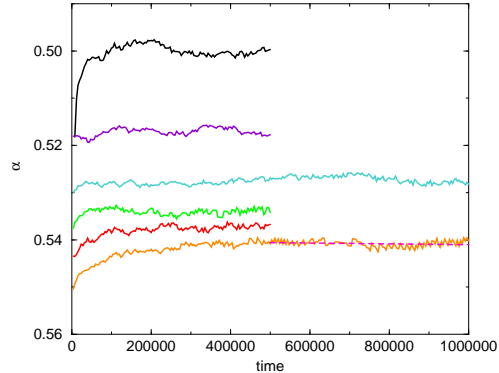}}
\caption{Local slopes of the kink density decay for symmetrical 
initial conditions $\rho_0(I1)=\rho_0(I2)=$
$0.3$,$0.2$,$0.15$,$0.1$,$0.05$ (from bottom to top curves).
The simulation result for one species (ARW) is also shown (top curve)}
\label{deacysl}
\end{figure}

To check whether the observed deviation from standard ARW 
behaviour is only a cross-over effect, or it heralds some basic 
feature of altered kinetics we have performed very long-time 
($t_{max} = 10^6$ MCS) simulations on systems with $L=24000$ 
(Fig. \ref{decay}) \cite{ASP}.
Throughout the whole paper $t$ is measured
in units of Monte-Carlo sweeps.
Figure \ref{deacysl} shows the results of simulations.
It is seen that the deviation from the standard ARW value of the  
decay exponent remains present asymptotically as well:
the local slopes of the decay curves
\begin{equation}
\alpha(t) = {\ln \left[ \rho(t) / \rho(t/m) \right] \over \ln(m)} \label{slopes}
\end{equation}
(where we use $m=8$ usually) go to constant values.
{\em Moreover, another interesting feature has
become apparent: the kink-decay exponent depends on the initial
concentrations }of the components $\rho_0(I1)=\rho_0(I2)$  and in
such a way  that for higher initial kink density (lower average 
distance between the kinks) the decay is faster.
Asymptotically, as $\rho_{0}\to 0$, the average distance of dissimilar
kinks goes to zero
and the decay exponent tends to the ARW value: $\alpha\to 0.5$.

In the case of asymmetric initial condition  ($\rho_0(I1)\ne\rho_0(I2)$)
$K1$-s and $K2$-s decay with different rates. The type that
has smaller initial density decays faster. Example: in case of
$\rho_0(I1)=1/9$, $\rho_0(I2)=1/3$, $K2$ decays roughly like $t^{-0.5}$
(unperturbed by $K1$-s) but the local slopes of the log-log $\rho(K1) - t$
dependence decrease from $-0.5$ strongly.

\section{Simulations from an active seed}\label{seedsim}

The cluster simulations \cite{GrasTor} were started from a state
with uniformly distributed $A$-s and $I1$-s except a single $I2$
pair in the middle and the following characteristic quantities
for the  $I2$-s
were followed:
\begin{itemize}
\item the average number of $I2$-s, $N_{I2}(t)$,
\item their survival probability $P_{I2}(t)$,
\item and the average mean square distance of spreading of $I2$'s
from the center $R^2_{I2}(t)$
\end{itemize}
The above quantities were averaged over $N_s$ independent runs
at the CDP2 point ( in case of $R^2_{I2}(t)$ only for surviving
samples ).
At the critical point we expect these quantities to behave
for $t\to\infty$, as
\begin{equation}
N_{I2}(t)\propto t^{\eta} \ , \label{state}
\end{equation}
\begin{equation}
P_{I2}(t)\propto t^{-\delta} \ , \label{statd}
\end{equation}
\begin{equation}
R^2_{I2}(t)\propto t^z \ . \label{statr}
\end{equation}

\begin{figure}[h]
\epsfxsize=70mm
\centerline{\epsffile{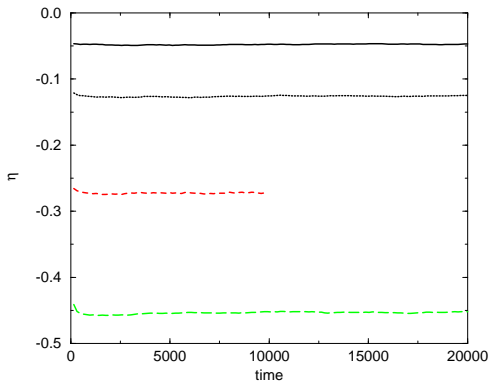}}
\vspace{2mm}
\caption{Local slopes of the number of $I2$-s. The initial state is uniformly
distributed with initial densities: $\rho_0(I1)=0.1$ (solid line),
$0.25$ (dotted line), $0.5$ (dashed line), $0.75$ (long-dashed line).
\label{seedI5_e}
}
\end{figure}

\begin{figure}[h]
\epsfxsize=70mm
\centerline{\epsffile{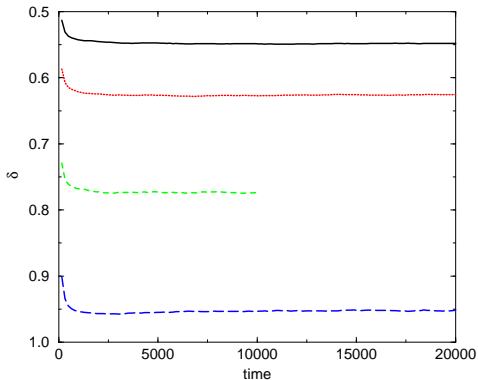}}
\vspace{2mm}
\caption{The same as on Fig. \ref{seedI5_e} for the cluster survival probability.
\label{seedI5_d}
}
\end{figure}

Upon varying the initial density $\rho_0(I1)$, for the exponents 
$\delta$ and $\eta$ (defined similarly to eq.(\ref{slopes}), the
local slopes of  $N_{I2}(t)$ and $P_ {I2}(t)$) 
continuously changing  values have been observed 
(Figs. \ref{seedI5_e}, \ref{seedI5_d}).
The deviation of these exponents from those of the single-species 
annihilation random walk process: $1/2$ and $0$, respectively is 
remarkable. 
The spreading exponent, $z$, on the other hand, seems to 
be constant within numerical accuracy and equals that of 
the single species annihilating random walk : $z=2/Z=1$ 
such that the generalised hyper-scaling law of the
compact directed percolation \cite{hyper2}
\begin{equation}
\eta+\delta=z/2 
\end{equation}
is satisfied.
In this respect it is important that $\eta$ has been found to be
negative.


\section{Cluster simulations of Compact directed percolation 
confined in a parabola. }\label{parabsim}

To understand the physics of our numerical results up to now
we set up a parallelism with an other case where DP
process is bounded by parabolic space-time boundary conditions. 
We perform simulations on the compact cluster version of this 
and compare the results with those of the GDK model in 
section \ref{summary}.

Kaiser and Turban have investigated \cite{K&T,K&T-2} the $1+1$
dimensional  DP process limited by a special, parabolic boundary 
condition in space and time directions:
\begin{equation}
y = \pm C t^k
\label{par}
\end{equation}
where $C$ changes under uniform length rescaling (by $b$) to:
\begin{equation}
C^{'}=b^{Zk-1}C
\label{c'}
\end{equation}
Here  $Z$ is the dynamical critical exponent.
By referring to conformal mapping of the parabola to straight lines
and showing it in the mean-field approximation Kaiser and Turban claim that
for $k<1/Z$ this surface gives relevant, for $k>1/Z$ irrelevant and for
$k=1/Z$ marginal perturbation to the DP process.
The marginal case results in $C$ dependent non-universal power-law decay,
(for details see next section), while for the relevant case
stretched exponential functions have been
obtained. The above authors have given support to  this claim
by numerical simulations.

We have investigated the effect of parabolic and reflecting boundary conditions
for the CDP2 process numerically. Time-dependent cluster
spreading simulations have been performed in the GDK model
with parabolic boundaries such that at each time step
the simulation region is bounded by two $I1$-s at
$y_{min}$ and $y_{max}$, where
\begin{equation}
y_{min} = L/2 - 2 - C \ t^k
\end{equation}
\begin{equation}
y_{max} = L/2 + 2 + C \ t^k \,
\end{equation}

\begin{figure}[h]
\epsfxsize=60mm
\centerline{\epsffile{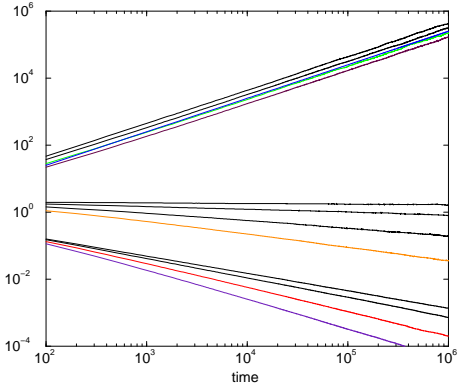}}
\vspace{2mm}
\caption{Parabola cluster confinement simulations for CDP.
Middle curves: $N_{I2}(t)$ ($C=2,1.5,1.2,1$ top to bottom);
Lower curves: $P_{I2}(t)$ ($C=2,1.5,1.2,1$ top to bottom);
Upper curves: $R^2_{I2}(t)$ ($C=2,1.5,1.2,1$ top to bottom).
\label{parab}
}
\end{figure}

\begin{figure}[h]
\epsfxsize=70mm
\centerline{\epsffile{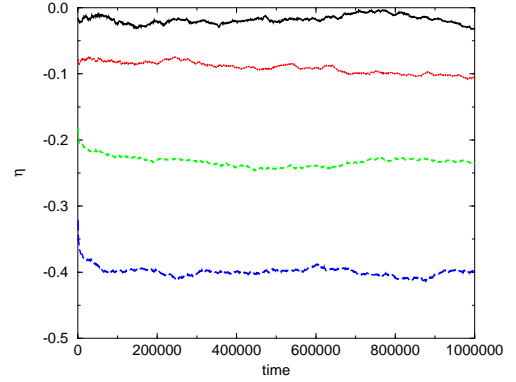}}
\vspace{2mm}
\caption{Parabola cluster confinement simulations for CDP.
Local slopes of $N_{I2}$ for different values of $C$.
Solid line: $C=2$, dotted line: $C=1.5$, dashed line: $C=1.2$,
long-dashed line: $C=1$.
\label{parab_e}
}
\end{figure}

\begin{figure}[h]
\epsfxsize=70mm
\centerline{\epsffile{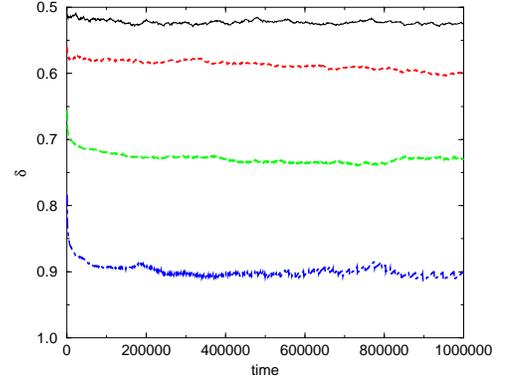}}
\vspace{2mm}
\caption{The same as Fig. \ref{parab_e} for the cluster survival
probability.
\label{parab_d}
}
\end{figure}

\begin{figure}[h]
\epsfxsize=70mm
\centerline{\epsffile{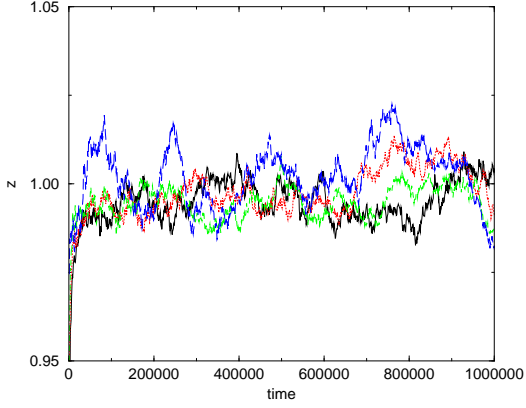}}
\vspace{2mm}
\caption{The same as Fig. \ref{parab_e} for the exponent of the
cluster spreading $R^2_{I2}$.
\label{parab_z}
}
\end{figure}

Two $I2$-s have been put initially at the centre $(L/2, L/2+1)$ and some initial
space (two $A$-s to the left and right) between $K1$-s and $K2$-s has been added.
Therefore the role of $I1$-s now is purely the formation of parabolic
boundaries around $I2$-s and in fact {\em we investigate the plain CDP process
with reflective boundary conditions}.
A typical 1+1 dimensional run looks like as shown below  (1, 2 and 0 stand for
$I1$, $I2$ and $A$, respectively):
\bigskip
\bigskip
\begin{verbatim}

                        <-- y -->

        |               10022001
        |               10222001
                       1002200001
        t              1022000001
                       1222200001
        |              1022220001
        |              1022220001
        |             100222200001
        |             100222220001
        V             100222200001
                      100222000001
                      102222000001


\end{verbatim}
When we fixed the exponent at  $k=1/2$, to make the situation marginal
we found continuously changing exponents for the exponents of the 
survival probability $P_{I2}$ and the number of $I2$-s,$N_{I2}$,  
by varying the shape $C$ (Fig. \ref{parab}).
One can see that the exponent-slopes of $N_{I2}(t)$ (Fig. \ref{parab_e})
and those of $P_{I2}(t)$ (Fig. \ref{parab_d}) change by varying $C$.
The spreading exponent of the $I2$-s, $z$, seems to be constant: equal
to unity.
(Fig. \ref{parab_z}).
These results are very similar to those of the seed simulations in GDK
of section \ref{seedsim}.
The analysis based on local slopes  (Figs. \ref{parab_e}, \ref{parab_d},
\ref{parab_z})
shows again plateaus for high values of $t$, indicating true 
power-law behaviours.
The magnitude of the exponent characterizing the decay of the density
of $I2$-s decreases as $C$ is increased
reminiscent of a similar situation in  \cite{K&T}.

The survival exponent changes in such a way that the hyper-scaling
relation valid in case of compact directed percolation\cite{hyper2}:
$${z/2}=\eta + \delta = 1/2$$ is fulfilled. In this case it is important, 
again, that $\eta$ takes negative values.

\section{Theoretical considerations for CDP confined in a parabola}
\label{parabMF}
\subsection{Anisotropic scaling}

In 1+1 dimensional anisotropic systems the correlation length
diverges as $\xi_\parallel\!\sim\! t^{-\nu_\parallel}$ in time
and as $\xi_\perp\!\sim\! t^{-\nu}$ in the
space direction with a dynamical exponent
$Z\!=\!\nu_\parallel/\nu$ ($\nu$ is also denoted as $\nu_\bot$ in the literature). 
Covariance under a change of the length
scales then requires two different scaling factors,
$b_\parallel\!=\! b^Z$ and $b_\perp\!=\! b$~.  

We  consider now a system displaying anisotropic critical behaviour and
limited by a free surface in the  $(t,y)$ plane as given in eq(\ref{par}).
Under rescaling, with $t'\!=\! t/b^Z$ and $y'\!=\! y/b$, $C$ transforms 
according~to eq.(\ref{c'}),as discussed in the previous section.

In the marginal case, which we will consider now,
$Z\!=\!1/k$, the scaling dimension $x_m$ of the tip order
parameter becomes $C$--dependent ${x_m(C)}$.

The order parameter correlation function between the origin and a
point at $(t,y)$ transforms as
\begin{equation}
G\left(\Delta,t,y,{1\over C}\right)=
b^{-2x_m}G\left(b^{1/\nu}\Delta,{t\over b^Z},
\frac{y}{b},{b^{1-Zk}\over C}\right)
\label{G1}
\end{equation}
when $L$ is infinite. With $b\!=\! t^{1/Z}$, equation (\ref{G1}) leads to:
\begin{equation}
G\left(\Delta,t,y,{1\over C}\right)=t^{-2x_m/Z}g\left({t\over
\Delta^{-\nu_\parallel}},{y^Z\over t},{t\over l_C}\right).
\label{G2}
\end{equation}
Here $l_C=C^{Z\over{1-Zk}}$, $\Delta=\frac{(p-p_c)}{p_c}$ and
$x_m$ is the scaling dimension of the order parameter. The latter is
 connected to
$\beta$, the critical exponent of the order parameter via $\beta=\nu x_m$.
We will use this scaling form in the following.
$\beta $ is the usual order-parameter exponent,
defined, for the DKCA, through $\rho_1 \propto
(p - p_c)^{\beta}$, for $p > p_c$; $\rho_1$ is the stationary density of 1's.
In case of a first order transition as is the case with compact
directed percolation the following considerations hold.

As already mentioned,
\begin{equation}
P(t) \propto t^{-\delta},
\end{equation}
is the  survival probability of $1's$
for spread of particles (1's, in our notation) about the origin.
Away from the critical point
$\beta' $ governs the ultimate survival probability (starting from a localized
source): $P_{\infty} \equiv \lim_{t \rightarrow \infty} P(t) \propto
{(p - p_c)}^{\beta'}$.
It is known that $\beta' = 1$ in CDP. \cite{DoKi}
The order-parameter exponent, $\beta$,
however, is {\em zero}.  This is because $p=1/2$
marks a {\em discontinuous} transition, by symmetry.
$\rho_1=0$ for $p < 1/2$ and
$\rho_1 = 1$ for
$p > 1/2$;  strictly speaking, $\beta$ is not defined here
 but it is natural to associate the value $\beta = 0$ with
the discontinuous transition.

This problem with the ill-defined exponent $\beta$ can be avoided
following the lines of Grassberger and de la Torre's scaling argument
\cite{GrasTor}
for discontinuous transitions as it has been presented by Dickman and
Tretyakov\cite{hyper2}.  Consider a model
with a transition from an absorbing to
an active state at $\Delta = 0$, with exponents $\delta$, $\eta$, $z$,
and $\beta'$
defined as above.
Suppose, however, that the order parameter $\rho$ is discontinuous, being
zero for $\Delta < 0$, and
\begin{equation}
\rho = \rho_0 + f(\Delta),
\end{equation}
for $\Delta > 0$, where $\rho_0 > 0$, and $f$ is continuous and vanishes
at $\Delta = 0$.  According to the scaling hypothesis for 
spreading from a source
there exist  two scaling functions, defined {\em via}
\cite{GrasTor}

\begin{equation}
\rho(y,t) \sim t^{\eta - dz/2}\tilde G(y^{2}/t^{z}, \Delta t^{1/\nu_{||}}),
\end{equation}
and
\begin{equation}
P(t) \sim t^{-\delta} \Phi(\Delta t^{1/\nu_{||}}).
\end{equation}
(Here $\rho(y,t)$ is the local order-parameter density.
 $\tau \sim \Delta^{-\nu_{||}}$.)  Existence of the limit
$P_{\infty}$
implies that $\Phi (x) \sim x^{\beta'} $ as $x \rightarrow \infty$,
with $\beta' = \delta \nu_{||}$.  In a surviving trial, the local
density must approach the stationary density $\rho$ as $t \rightarrow \infty$,
so $\rho(y,t) \sim \Delta^{\beta'} \rho_0 $,
for $t \rightarrow \infty$ with fixed $y$, and $\Delta $ small but positive.
It follows that $\tilde G(0,x) \sim x^{\beta'} $ for large $x$.

An important consequence is that
we can use as scaling dimension of the order parameter for CDP the value
$\beta'$ in the relation $x_m={\beta\over\nu}$ instead of $\beta$.
Via scaling relations $\beta'=\delta\nu_{||}$, the values obtained by
computer simulations for $\delta$ will be compared with results for
CDP+parabolic boundary conditions. In this context the connection,
again via scaling relation, between $\delta$ and the decay exponent
of the density of kinks when starting from a random initial
state $\alpha$ will also be made use of .

\subsection{Mean field analysis for CDP confined in a parabola}

In this section we will follow the lines of the mean-field analysis of the
$1+1$ dimensional DP process confined by a parabola
as given in ref.\cite{K&T-2}, but now applied to compact directed percolation.

The basic processes are:
\begin{figure}[h]
\epsfxsize=70mm
\centerline{\epsffile{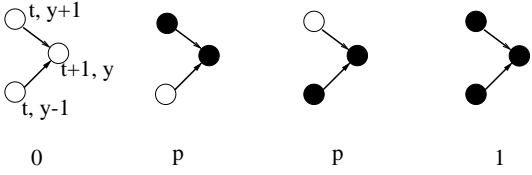}}
\vspace{2mm}
\caption{Site update rules for compact directed percolation.
\label{cperc}
}
\end{figure}
The order parameter correlation function is the probability density $P(t,y)$
for a site at $(t,y)$ to be connected to the origin.

First we consider the case without confinement. In mean-field
approximation one can set up an equation for the connectedness at $(t+1,y)$:
\begin{eqnarray}
P(t+1,y) &=& p \{ P(t,y+1) [ 1-P(t,y-1) ] + \nonumber\\
P(t,y-1) &[& 1-P(t,y+1) ] \} + P(t,y+1) P(t,y-1)
\end{eqnarray}
Going to the continuum limit the following differential equation is
obtained:
\begin{equation}
\frac{\partial P}{\partial t} = p \frac{\partial^2 P}{\partial y^2} + (2p-1)P
+ (1-2p)P^2 \label{diff}
\end{equation}
The homogeneous, stationary solution of eq.(\ref{diff}) is:
\begin{eqnarray}
P_0 = \left\{
\begin{array}{lll}
1 & \mbox{for} & p > 1/2 \\
0 & \mbox{for} & p \le 1/2 
\end{array}
\right.
\end{eqnarray}
describing a first order transition for CDP at $p_c=1/2$, as it is the case.
At the transition, $p=p_c$, eq.(\ref{diff}) reduces to
\begin{equation}
\frac{\partial P}{\partial t} = \frac{1}{2}\frac{\partial^2 P}{\partial y^2}
\label{ARW2e}
\end{equation}

This is the ordinary diffusion equation with RW solution
\begin{equation}
P(t,y)= \frac{\exp\Bigl(-{y^2\over
2t}\Bigr)}{\sqrt{2\pi t}}
\label{diff0}
\end{equation}
which is exact in the CDP case. From comparison with the scaling 
form in the previous subsection,
the following (well-known) exponents for CDP arise:
\begin{equation}
\nu_\parallel=1\qquad\nu=1/2\qquad Z=2\qquad x_m={1\over 2}
\label{compact}
\end{equation}

On a parabolic system, we use the new variables $t$ and
$\zeta(t,y)\!=\! y/t^k$
for which the free surface is shifted to $\zeta\!=\!\pm C$ and equation 
(\ref{ARW2e}) is changed into
\begin{equation}
{\partial P\over\partial t}={1\over 2t^{2k}}{\partial^2P\over\partial
\zeta^2}+k{\zeta\over t}{\partial P\over\partial\zeta}
\label{diff1}
\end{equation}
with the boundary condition $P(t,\zeta\!=\!\pm C)\!=\!0$.
Through the change of function
\begin{equation}
P(t,\zeta)=\exp\big[-{k\over 2}\zeta^2t^{2k-1}\big]Q(t,\zeta)
\label{diff2}
\end{equation}
equation (\ref{diff1}) leads to
\begin{equation}
{\partial Q\over\partial t}={1\over 2t^{2k}}{\partial^2 Q\over\partial
\zeta^2}+{k\over 2}\Bigl[(k-1)\zeta^2t^{2k-2}-{1\over t}\Bigr]Q
\label{diff3}
\end{equation}
for which the variables separate when $k\!=\!1,1/2$ or $0$.
These values of $k$ just correspond
to irrelevant, marginal and relevant perturbations.

For $k\!=\!1$ the critical
behaviour is the same as for un--confined percolation as expected for an
irrelevant perturbation.

For the true parabola which is the marginal geometry, one may use equation
(\ref{diff1})
with $k\!=\!1/2$ to obtain
\begin{equation}
t{\partial P\over\partial t}={1\over 2}{\partial^2 P\over\partial
\zeta^2}+{\zeta\over 2}{\partial P\over\partial\zeta}\qquad \zeta={y\over
t^{1/2}}
\label{diff6}
\end{equation}
which is of the form studied in \cite{Turban92} for the directed walk problem.
Writing
$Q(t,\zeta)\!=\!\phi(t)\psi(\zeta)$ leads to the following eigenvalue 
problem for
$\psi(\zeta)$
\begin{equation}
{1\over 2}{{d^2}\psi\over{d{\zeta^2}}}+{\zeta\over
2}{d\psi\over{d\zeta}}=-\lambda^2\psi
\label{diff7}
\end{equation}
with $\phi(t)\!\sim\! t^{-\lambda^2}$.
 The solution is
obtained as the eigenvalue expansion
\begin{equation}
P(t,y)=\sum_{n=0}^\infty B_n\ t^{-\lambda_n^2}\
_1F_1\Bigl[\lambda_n^2,{1\over 2};-{\frac{y^2}{2t}}\Bigr] .
\label{diff8}
\end{equation}
The behaviour at large $t$ is governed by the first term in this
expansion which decays as $t^{-\lambda_0^2}$, i. e. with a $C$--dependent
exponent as expected for a marginal perturbation. {\it The dimension of the
tip--to--bulk correlation function is the sum of the tip and bulk
order parameter dimensions,  the first one being variable.} Comparing with the
form of the decay in eq.(\ref{G2}) gives $\lambda_0^2\!={\![x_m^{mf}(C)\!+\!
x_m]}/Z$ and, using (\ref{compact}), the tip order
parameter dimension is given by
\begin{equation}
x_m^{mf}(C)=2\lambda_0^2-{1\over 2} .
\label{diff9}
\end{equation}
Its dependence on $C$ is shown in figure~2. of \cite{K&T}

Analytical results can be obtained only in limiting cases which have
already been
discussed in \cite{Turban92}. When $C$ is infinite, $\lambda_0^2\!=\!1/2$,
only the first
term in the expansion remains, which satisfies the initial and boundary
conditions, giving back the free solution in equation (\ref{diff0}).
For large C-values the tip exponent is
$x_m^{mf}(C)={1\over 2}+\sqrt{2\over\pi} C \exp\Bigl(-{C^2\over
2}\Bigr)[1+O(\varepsilon)] $
where $\varepsilon$ is the correction term itself. For narrow systems, the
hyper-geometric function gives a cosine to leading order in $C^2$. One obtains
the following asymptotic behaviour in $t$
\begin{equation}
P(t,y)\sim t^{-\pi^2/8C^2}\cos\left({\pi y\over 2C\sqrt{t}}\right)
\label{diff10}
\end{equation}
and the tip exponent diverges as $\pi^2/4C^2$.

For $0<k<1/2$ the dependence on $t$ is expected to be a stretched
exponential function. For details see \cite{K&T,Turban92}.

\section{Annihilating random walk of two species with exclusion}
\label{ARW2e-sect}

To check our results concerning the scaling properties of  kinks
in the GDK model at the CDP2 point
we have carried out an explicit simulation of the annihilating
random walk of two species ($A$, $B$) with
exclusion. The model we have investigated has been suggested by
Hinrichsen \cite{Hpriv} and
is as  follows. A(B) will hop to a neighbouring empty site with probability
$p1A$ ($p1B$) or annihilate with a neighbour A (B) with probability
$p2A$ ($p2B$) while A and B do not react when getting into neighbouring
positions.
The initial configuration  was chosen in such a way that
allows  pairs of the same kind to annihilate always within some finite
time interval (i.e. the
system evolves into an empty state),namely:
\begin{verbatim}
       ... A.A.B.B.A.A.A.A.B.B....
\end{verbatim}
That means that $AA$ and $BB$ pairs have been put in a 1d ring
with initial probability $\rho(0)$. Had we not chosen the initial state
like this the system would have ended up in some finite particle 
configuration where $A$-s and $B$-s
follow each other alternatingly, separated by arbitrary empty regions.
(This initial configuration is in agreement with the arrangement of the
two kinds of kinks in some random initial state of the GDK model, too. )
The  probabilities $p1A$, $p1B$, $p2A$ and $p2B$
have been chosen to be unity, to achieve maximum simulation effectiveness;
no qualitative difference in the results have been found upon lowering
them.

Clearly this process is different from the simple annihilating
random walk of two species $A+B\to\emptyset$ \cite{LeeAB},
therefore we may expect that a field theory describing this
model (which, however, is still missing) would result in a
different fixed point with different critical exponents as well.
Furthermore one can argue that when comparing the simple random walk
and the random walk+exclusion (SEP) processes one also observes
different dynamical behaviours.
This latter case is nothing else but the $T=0$ dynamics of
the $1d$ Ising Model with Kawasaki exchange, where we have different
domain growth properties than in case of a simple random walk.

\begin{figure}[h]
\epsfxsize=70mm
\centerline{\epsffile{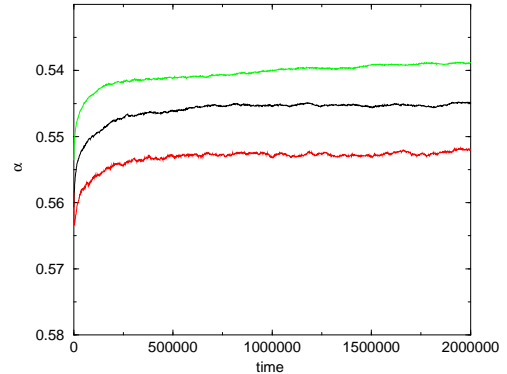}}
\caption{Local slopes of particle decay in the annihilating random walk
+ exclusion process of two species results ($L=24000$).
The initial conditions are: $\rho_0 = 0.1, 0.25, 0.5$ from bottom to top curve.
\label{arw2}
}
\end{figure}

An extensive numerical simulation with look-up table algorithm seems to confirm
this expectation. As Figure \ref{arw2} shows the slopes of the density decay
started from the special pairwise random states described above
depend on the initial density $\rho(0)$.

The local slopes tend to constant values greater than $\alpha=0.5$ in agreement
with the GDK kink results. The level-off in case of $\rho_0=0.5$ happens only
for $t > 1.5\times 10^6$ MCS. 
The average $AA$ and $BB$ distances confining an other type of particle
have also been measured during the simulations, that enables us to extract 
the amplitude ($C$) of the confinement in the function fitted 
($C\times t^{-\alpha}$). These values will be used to compare the
results with those of the GDK (see next section).

\section{Summary of time dependent results}\label{summary}

Since in all of the previously shown cases we found non-universal scaling
depending on the initial conditions and the generic model to account for such
behavior seems to be the CDP2 with parabolic boundary condition,
we have decided to measure
the region of confinement in all cases and plotted the survival probability
exponents $\delta$ and the kink decay exponents $\alpha$ as a function of the
shape of the measured parabola.

In the present  case $ \beta^,$
the final survival probability of a cluster plays the role of the
order parameter exponent $\beta$ as explained at the end of Sec.VI
and for the characteristic exponents we have:
$\delta=\beta^{'}/{\nu_{\perp}}$. Thus we have plotted the results 
for $\delta(C)$. In a common graph the fitted values for $\alpha(C)$ are
also shown; on the level of kinks the order parameter $\beta$ is
connected to $\alpha$ in the same way as $\beta'$ to $\delta$ for
'spins' (see eg. \cite{ujMeOd}).

\begin{figure}[h]
\epsfxsize=70mm
\centerline{\epsffile{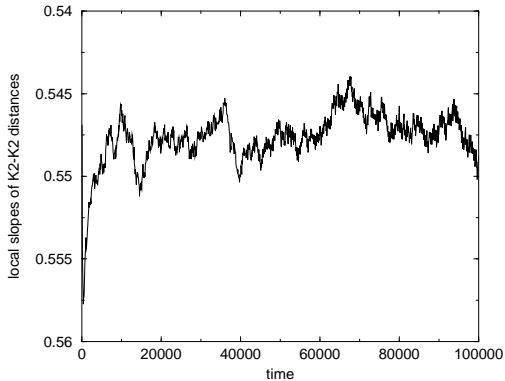}}
\vspace{2mm}
\caption{Local slopes of the $K2-K2$ neighbour distances in GDK model
of size $L=24000$. The initial state is uniform with 
$\rho_0(I1)=\rho_0(I2)=1/3$
\label{K2-K2}
}
\end{figure}

For random initial conditions in the GDK model the characteristic distance
between two neighbouring kinks of a given type has also been measured.
The average neighbour distance $l_{K2-K2}$ shown on Fig. \ref{K2-K2} have 
been obtained for initial densities: $\rho_0(I1)=\rho_0(I2)=1/3$.
The power law increase for large $t$ (see the plateau for $t > 30.000$)
with the same scaling exponent as the decay exponent is
not very surprising because $\rho_{kink}(t) \propto 1/l_{K2-K2}$.

Since the $K2-K2$ and the $K1-K1$ pairs confine the motion of each other
( a $K1 - K2$ pair can not exchange to $K2 - K1$) this power-law
increasing length scale imposes a 'stochastic' boundary condition
(pressure on kinks) with a mean value of a parabola similarly that was
investigated by Kaiser and Turban \cite{K&T} in case of $1+1$ d DP processes
\cite{K&T-2} and adapted for the case of a CDP-like first order transition in
section VI. As discussed before, the scaling dimension of the order parameter
changes continuously with the amplitude of the parabolically growing confining
box size if it grows with the same exponent as the cluster inside.

\begin{figure}[h]
\epsfxsize=70mm
\centerline{\epsffile{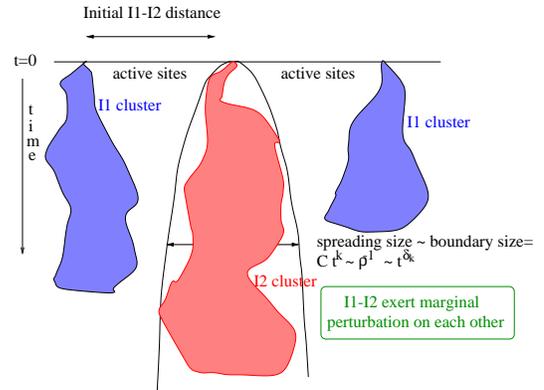}}
\caption{Possible explanation for the non-universal scaling.
The symmetric $I1$ an $I2$ clusters exert marginal exclusion
perturbation on each other.
\label{expl}
}
\end{figure}

In our case we encounter similar situation. The kink density decay exponent
$\alpha$ seems to vary continuously in case of symmetrical initial
conditions. The initial conditions effect the amplitudes of the density
decays (as was shown to be valid by field theory for pure the reaction
diffusion of A,B particles \cite{LeeAB}) and therefore the amplitudes of
the confinement region sizes ($C$) (see Fig. \ref{expl}).
To compare our results with those of \cite{K&T,K&T-2} the form
$A + C \ t^{\alpha}$ has been fitted to the $l_{K2-K2}(t)$
distances determined from the density decay simulations
(assuming $l_{K2-K2}(t) = 2/\rho_{kink}(t)$).
The following table summarises the results for GDK with random initial
conditions:

\begin{center}
\begin{tabular}{||c|c|c||}
\hline\hline
$\rho_0$  & $C$      & $\alpha$\\ \hline
$0.0$    &  $\infty$ & 0.5000(3)\\ \hline
$0.05$   & $14.09$ & $0.517(2)$\\
$0.10$   & $7.62$  & $0.528(2)$\\
$0.15$   & $6.11$  & $0.534(2)$\\
$0.2$    & $5.85$  & $0.537(2)$\\
$0.3$    & $4.18$  & $0.540(1)$\\
\hline\hline
\end{tabular}
\end{center}

This is in agreement with Fig. 8 of \cite{K&T-2},
where an increasing $C$ causes a decreasing exponent.
Note that the first line in the table corresponds to the simple ARW
process, therefore there is no confinement (amplitude $C$ is $\infty$).

{\sl The main difference between the rigid parabola boundary case and
the "stochastic" confinement is that in the latter case the boundary
generates an additional noise to the motion of confined particles.
Therefore we don't simple have "free" particles confined in a parabola, but
they are also perturbed by the noise in such a $K1\leftrightarrow K2$
symmetrical way that the outcome perturbation is marginal.}

\bigskip

In ARW simulations again the form $A + C \ t^{\alpha}$ has been fitted 
for the measured $AA$, $BB$ distances.

In cluster simulations we fitted the form: $y=C\times t^{\delta}$ 
for the region of confinements and determined the respective $C$-s 
in all cases.

\begin{figure}[h]
\epsfxsize=70mm
\centerline{\epsffile{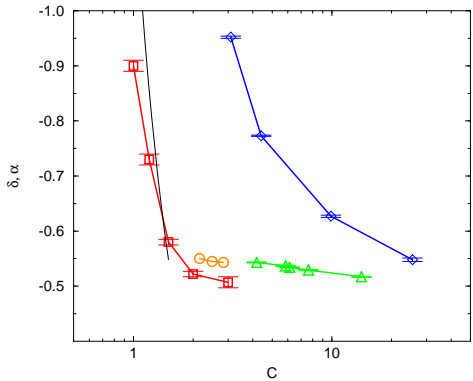}}
\caption{Confinement shape parameter ($C$) dependence of 
exponent $\delta$ for cluster simulations in parabola (squares), 
$I2$ cluster (diamonds) and $\alpha$ in simulations from 
uniform initial conditions in ARW2e (circles), 
GDK (triangles). The solid line shows the 
$\pi^2/8 C^2$ mean-field approximation.
\label{C-d}
}
\end{figure}

As Figure \ref{C-d} shows we obtained similar monotonically decreasing
curves in all cases that also agrees with the results of \cite{K&T}.
The GDK-uniform and the ARW2 results seem to lie on the same curve.
The spreading simulation results of GDK are different from those of the
CDP+parabolic boundary condition case.

This can be understood, however, since in the former
case the confined particles ($I2$-s) have a back effect on the bulk ($I1$-s)
particles, while this is not the case when the boundary is fixed.

We also show the $\pi^2/{8C^2}$ curve, determined as the asymptotic solution
$C\to 0$ of the mean-field approximation, see section VI eq.(\ref{diff10}).
This seems to be in fair agreement with the case of CDP+fixed parabolic 
boundary condition.

\section{Generalisation for $N>2$: symmetric annihilating exclusion process
of $N$ species} \label{Gener}

We have carried out preliminary simulations in the generalized
version of the model introduced in section VII. The system
was started  from
configurations like:
\begin{verbatim}
        .... A..A.B...B..C.C..DD....E..EFF ...
\end{verbatim}
where species of the same type can annihilate each other but different types
can not exchange. Our results show that concerning the time dependence of
the density
 the deviations from the square root  decay persist
for $N>2$ and this property  seems to remain valid also for 
$N\to\infty$.

\begin{figure}[h]
\epsfxsize=70mm
\centerline{\epsffile{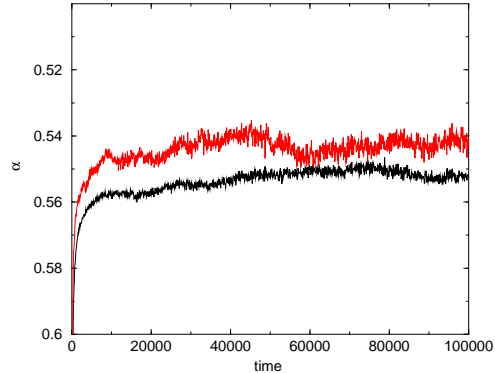}}
\vspace{2mm}
\caption{Results for density decay in symmetric annihilating
exclusion process of $N=3,4$ (top, bottom curve) species. 
\label{fig16.eps}
}
\end{figure}

 Fig. \ref{fig16.eps} shows that a similar level-off can be observed in the
local slopes as in the $N=2$ case with $\alpha > 0.5$.  
A tendency of increasing $\alpha$ with increasing $N$ is apparent in our
simulations; the growing 
effect of finite size corrections, however, prevented us from going
further, for higher values of $N$, in this study.

\section{GDK on the line of compactness} \label{pneq}

On the line $q=1$ the role of the absorbing states
($I1$ and $I2$) is symmetric. The above reported simulation results
for the GDK model refer to  $p=1/2$, the CDP2 point. Now we will discuss
the situation for $p\neq 1/2$.
For $p>1/2$ the creation of  new $A$-s happens
with probability greater than 0.5, the active domain size grows exponentially
and the inactive regions die out quickly (and symmetrically);
the all-A- phase plays the same role as the all- 0- phase of the original
DK automaton \cite{DoKi}.
 The deviation from the DK picture is quite apparent,however,
  for $p<1/2$, as instead of the all-1-phase of DK, for all values of
  $p$ with the exception of $p=0$,  Glauber-Ising -like kinetics
  governs the motion of kinks. The kinks here are extended objects (A-s),
  somewhat similarly to those in Grassberger's CA models\cite{gra84}( for
  $p<p_c$, where $p_c$ is the critical point of a parity conserving phase
  transition )
  where also kinks of
  different extension (and there even of different structure) separate
  absorbing-phase clusters. (At $p=0$ diffusion stops and a striped
  space-time picture of I1 and I2 domains freezes in, again like in
  Grassberger's model A cited above.)
   The average size of $A$-s goes to zero between two domains of the same type
quickly, the kink 'particles' of same type perform a {\bf biased} random walk
toward each other. On the other hand between domains of different types remains
a film of $A$-s of average size $1$, since a collision of an $I1$ and $I2$
domain always creates a new $A$ at the next time step.
That means that kinks of different types still block the motion of each other.
Therefore the role of $A$-s is similar to the kinks of the $T=0$ Glauber Ising
model.
On the whole line of $1/2>p>0$, in the long time limit, therefore one
can expect the  number of such kinks to decrease as
($t^{-1/2}$). On the other hand this is a line of compactness, as
all clusters growing from a seed are compact, the characteristic
exponents ($\eta$, $\delta$ and $z$) though strongly dependent on $p$
and the composition of the initial state, satisfy the hyper-scaling
law valid for compact cluster \cite{hyper2}. This statement involves
that this line is a line of first order transition points with
order parameter exponent $\beta_s=0$. This first order transition
occurs in a symmetry-breaking ('magnetic') field coupled to the
$I1$ and $I2$ spins. For the detailed description of a similar situation
see \cite{MeOd98}.
All these features have been supported by simulations. As an example
we can give some results obtained at $p=0.4$ starting with a single
$I2$ in the sea of $I1$, $A$-s ($25\%$ of $I1$, $75\%$ of $A$ );
$\delta=0.45$, $z/2=0.47$, $\eta_{I2}= -0.02$.
It is worth mentioning, that for hyper-scaling to hold it is
again important that $\eta_{I2}$ is negative.

\section{Discussion}\label{DISCUS}

We have investigated numerically the one dimensional generalized 
Domany-Kinzel cellular automaton on a line  in
the plane of its parameters
where only compact clusters grow.
The two types of kinks in the simplest version of this compact GDK
model (two absorbing phases) follow
annihilating random walk  with exclusion (no reaction) between different 
types. 
The equivalence with an explicit two-species 
ARW model with exclusion  is shown
provided the initial state is prepared in such a way that the 
kinks are arranged in pairs with some density. High precision simulations
revealed that 
this system relaxes in a non-trivial way: the decay exponent of the kink
density depends on the
initial density of kinks. 
We argue that this is a kind of (internal) surface effect;
similar to the ARW process confined by a rigid
space-time parabola provided  the power of the parabola is chosen 
to be marginal.
This  case has been explicitly investigated with the result that
 the spreading 
exponents behave qualitatively the same way as expected from the
corresponding mean-field 
approximation. 
We have no proof of the marginality for the theory including
fluctuations, but rely on symmetry arguments.
If we assumed that particles would exert relevant perturbation 
($k<1/2$) on each
other, the corresponding parabola picture  would predict  stretched
exponential decay ( a behaviour that is very difficult to 
differentiate from power-laws 
by simulations) and  the local slopes should go to some higher value as a
function of time (meaning faster that any power-law decay). However our high
precision data show just the opposite, the local slopes {\it decrease}  
as a function
of time
tending to a value somewhat greater  than $1/2$. 

Nevertheless, the possibility of pure square-root decay masked by some
tremendously long crossover function can not be ruled out.
One could still expect a non-universal 
scaling of the survival probability of particles in the same way
as was observed in 
\cite{K&T,K&T-2} or in an other similar situation 
\cite{Krapi} where a diffusing "prisoner" 
confined by marginally growing cage was investigated.
In the latter case the the boundary condition was absorbing and an exact 
solution was possible giving an exponent for the survival probability 
which is a continuous function of the amplitude of the marginal parabola.
Furthermore the survival of a diffusing prisoner (with diffusivity D) 
inside a cage where both walls diffuse (with diffusivity A) has been
solved exactly and the decay exponent was found to be 
$\pi/2 \cos^{-1}(D/(D+A)$ \cite{benavra}.

Non-standard scaling in a 1d ARW model was also observed by Frachebourg 
et al. \cite{redner}. They have shown that the survival probability of 
particles in an ARW with one free boundary depends on the location of 
the particles. If we count the particles from the free boundary the 
survival probability of odd particles decay with exponent $0.225$, 
while those of even numbered decay with exponent $0.865$. 
The explanation for this is based on the fact that even numbered 
particles always have left and right neighbours during the process, while 
odd numbered particles lack one of the neighbours and since the ARW in 
1d is diffusion limited they can escape.
One can notice a similarity of this mechanism to the one in ARW2 models 
we investigated. Namely in our case there are infinitely many internal 
boundaries (generated by particles of different types which can not 
exchange sites).

 Recently Bray \cite{Bray99} has shown that the relaxation towards the
 critical state in the 2d XY model depends on the initial state. This
 is very different from what is expected from field-theoretical RG
 predictions that can not take into account low-dimensional topological
 effects. Moreover, Bray has shown that the non-universal behavior of
 the persistence exponent in this case can be described by the random walk of
 a particle moving under an attractive central  power-law force that
 creates marginal perturbation as compared to free random walk\cite{Bray99}
 This scenario is similar to ours since we also have particles with RW
 exhibiting pressure of marginal strength on each other.

Right after our submission two other preprints appeared on cond-mat
\cite{Satya,Kwon},
dealing with models very similar to those presented here and
reporting results which are in accord with ours for those quantities
they also investigated.

\vspace{3mm}
\noindent
{\bf Acknowledgements:}\\

The authors would like to thank Z. R\'acz, S. Redner, P. Arndt,
U. Tauber for useful remarks and H. Hinrichsen for taking part in 
the early stages of this work.
Support from Hungarian research fund OTKA (Nos. T-23791, T-25286 and 
T-23552) is acknowledged. One of us (N.M.) would like to thank
R. Graham for hospitality at the Fachbereich Physik of 
Universit\"at-GHS Essen, where this  work has been completed.
G. \'O. acknowledges support from Hungarian research fund 
B\'olyai (No. BO/00142/99) as well.
The simulations were performed partially on Aspex's System-V 
parallel processing system (www.aspex.co.uk).

\end{document}